\documentclass[12pt,notitlepage,a4paper]{article}
\newlength{\secskip}
\def\newsec#1{\vspace{\secskip}
\noindent
\makebox[\parindent]{{\bf #1. }}\ignorespaces
}
\usepackage{amsmath,amssymb,amscd,amsfonts}
\usepackage{mathtools}
\usepackage{url}
\def\eq{\begin{equation}}
\def\en{\end{equation}}
\def\eqa{\begin{eqnarray}}
\def\ena{\end{eqnarray}}
\newcommand{\cD}{{\mathcal D }}
\newcommand{\cQ}{{\mathcal Q }}
\newcommand{\cB}{{\mathcal B }}
\newcommand{\cV}{{\mathcal V}}
\def\Integers{\mathbb{Z}}

\def\Reals{\mathbb{R}}
\def\Disk{\mathbb{D}}
\def\PB{\cB}
\def\PBM{\PB(M)}
\def\IM#1#2{I_{M}\expval{#1,#2}}
\def\expval#1{\langle \, #1 \,\rangle}
\DeclareMathOperator\Hom{Hom}
\DeclareMathOperator\Aut{Aut}

\DeclarePairedDelimiter\norm{\lVert}{\rVert}%
\def\integral{\mathrm{int}}
\def\sing{\mathrm{sing}}
\def\distr{\mathrm{distr}}
\def\flat{\mathrm{flat}}

\newcommand{\overbar}[1]{\mkern2mu\overbracket[0.25pt][-1pt]{\mkern-2mu#1\mkern-5mu}\mkern 5mu}

\def\Aut{\mathbf{Aut}}
\begin{document}
\author{Daniel Friedan\\
New High Energy Theory Center, Rutgers University\\
Natural Science Institute, University of Iceland}
\date{November 13, 2017}
\title{A new kind of quantum field theory
of $(n{-}1)$-dimensional defects in $2n$ dimensions}
\maketitle

I describe a project to open a new territory of quantum field theory
where the fields live not on a space-time manifold but on certain
complete metric spaces of $(n{-}1)$-dimensional objects (defects) in a
$2n$-dimensional space-time $M$.  These metric spaces
are  ``quasi Riemann surfaces'';
they are formally analogous to
Riemann surfaces.
Every construction of a 2d conformal field theory is to give an analogous 
construction of a cft on the quasi Riemann surfaces,
and thereby a cft on $M$.
The global symmetry group of the 2d cft becomes a local 
gauge symmetry.
Ordinary local quantum fields in space-time
are constructed by restricting to small objects.
The project is based on writing the free $n$-form in $2n$ dimensions as 
the 2d gaussian model on the quasi Riemann surfaces.

This note is a summary of the main points of \cite{Friedan:2016mvo}.
References can be found there.
A condensed version of this note will appear as
\cite{Friedan:ChicheleyHall}.
More expositions are collected at \cite{Friedan:webpage}.

\newsec{1}
Let $M$ be euclidean space-time: an oriented conformal 
$2n$-manifold, compact, without boundary.
When $n=1$, $M$ is a Riemann surface.
The basic examples are $M=S^{2n} = \Reals^{2n}\cup\{\infty\}$.
The Hodge $*$-operator acting on $n$-forms
is conformally invariant,
\eq
(*\omega)_{\nu_{1}\cdots\nu_{n}}(x)
=
\omega_{\mu_{1}\cdots\mu_{n}}(x)
\,\frac1{n!} \,
\epsilon^{\mu_{1}\cdots\mu_{n}}{}_{\nu_{1}\cdots\nu_{n}}(x)
\qquad
*^{2}=(-1)^{n}
\en
Nothing else is used of the conformal structure on $M$.

\newsec{2}
The physical objects are represented mathematically
as the integral $(n{-}1)$-currents in $M$,
as constructed in Geometric Measure Theory \cite{FedererFleming}.
A $k$-{\it current} $\xi$ in $M$ is a distribution on the smooth $k$-forms,
\eq
\omega\in\Omega_{k}(M)\mapsto 
\int_{\xi}\omega
=
\int_{M} 
\frac1{k!} \omega_{\mu_{1}\ldots\mu_{k}}(x)
\,\xi^{\mu_{1}\ldots\mu_{k}}(x) d^{2n}x 
\en
The boundary operator $\partial$ on currents is dual to the exterior 
derivative,
\eq
\int_{\partial\xi}\omega = \int_{\xi}d\omega
\qquad
\partial^{2}=0
\en
A $k$-simplex in $M$, $\sigma\colon\Delta^{k}\rightarrow M$,
is represented by the $k$-current $[\sigma]$
which is the delta-function concentrated on 
$\sigma(\Delta^{k})\subset M$,
\eq
\sigma\colon\Delta^{k}\rightarrow M
\qquad
\int_{[\sigma]} \omega = \int_{\Delta^{k}} \sigma^{*}\omega
\en
A singular $k$-chain in $M$ is an
integer linear combination
of $k$-simplices in $M$,
$\sigma=\sum_{i}m_{i}\sigma_{i}$.
The {\it singular $k$-currents} $\cD^{\sing}_{k}(M)$
are the currents $[\sigma]=\sum_{i}m_{i}[\sigma_{i}]$
that represent the singular $k$-chains.
Examples are the $k$-submanifolds.
The current $[\sigma]$ represents the physical object in $M$ independent of
its expression as a combination of simplices.

The physical difference between two singular $k$-currents
is measured by
the {\it flat metric} $\norm{\xi_{1}-\xi_{2}}_{\flat}$,
\eq
\norm{\xi}_{\flat} = \inf\{\text{vol}_{k}(\xi-\partial \xi')
+\text{vol}_{k+1}(\xi')
\colon
\xi'\in \cD^{\sing}_{k+1}(M)
\}
\en
The space  of {\it integral $k$-currents} $\cD^{\integral}_{k}(M)$ is the metric completion of 
$\cD^{\sing}_{k}(M)$,
\eq
\cD^{\sing}_{k}(M) \subset \cD^{\integral}_{k}(M) \subset
\cD^{\distr}_{k}(M)
\qquad
\cD^{\integral}_{k}(M)\xrightarrow{\partial}\cD^{\integral}_{k-1}(M)
\en
The boundary of an integral current is an integral current.
$\cD^{\integral}_{k}(M)$ is a metric abelian group --- a 
complete metric space
and an abelian group.

\newsec{3}
Recall the 2d gaussian model, the free 1-form cft in 2d.
$j(x)$ is a 1-form
on a Riemann surface
satisfying
\eq
dj = 0
\qquad
d(*j)=0
\en
The integrals of $j$ and $*j$
are 0-forms  $\phi$, $\phi^{*}$
which take values in dual circles,
\eq
d\phi = j
\quad
d\phi^{*} = {*}j
\quad
\phi(x) \in \Reals/2\pi R\Integers
\quad
\phi^{*}(x) \in \Reals/2\pi R^{*}\Integers
\quad
R R^{*} =1
\en
$\phi$, $\phi^{*}$ are determined up to $U(1){\times} U(1)$ global symmetries
\eq
\phi(x)\rightarrow \phi(x)+a
\qquad
\phi^{*}(x)\rightarrow \phi^{*}(x)+a^{*}
\en
The vertex operator $V_{p,p^{*}}(x)$
describes a point defect
of charges $p,p^{*}$,
\eq
V_{p,p^{*}}(x)= e^{ip\phi(x) + 
ip^{*}\phi^{*}(x)}
\quad
p,p^{*}\in \frac1R \Integers \times\frac1{R^{*}} \Integers
\quad
V_{p,p^{*}} \rightarrow V_{p,p^{*}}\, e^{ipa + 
ip^{*}a^{*}}
\en

\newsec{4}
Recall the free $n$-form cft in $2n$ dimensions.
$F(x)$ is an $n$-form on the $2n$-manifold $M$ satisfying
\eq
dF = 0 \qquad d(*F) = 0
\en
The integrals of $F$ and $*F$
are $(n{-}1)$-forms  $A$, $A^{*}$ on $M$,
\eq
dA = F
\qquad
dA^{*} = *F
\en
which take values in dual circles
in the sense that
\eq
\int_{\xi} A \in \Reals/2\pi R\Integers
\quad
\int_{\xi} A^{*} \in \Reals/2\pi R^{*}\Integers
\quad
\forall\xi \in \cD^{\sing}_{n-1}(M)
\qquad
R R^{*} =1
\en
$A$, $A^{*}$ are
determined up to $U(1){\times} U(1)$ local gauge symmetries 
given by $(n{-}2)$-forms $f$, $f^{*}$
\eq
A\rightarrow A+df
\qquad
A^{*}\rightarrow A^{*}+df^{*}
\en
$(n{-}1)$-dimensional defects are 
described by fields $V_{p,p^{*}}(\xi)$ on $\cD^{\sing}_{n-1}(M)$,
\eq
\begin{gathered}
V_{p,p^{*}}(\xi) = e^{ip\phi(\xi) +ip^{*}\phi^{*}(\xi)}
\qquad
p,p^{*}\in \frac1R \Integers \times\frac1{R^{*}} \Integers
\\[1ex]
\phi(\xi) = \int_{\xi} A
\qquad
\phi^{*}(\xi) = \int_{\xi} A^{*}
\qquad
\xi \in  \cD^{\sing}_{n-1}(M)
\end{gathered}
\en
transforming by
\eq
\begin{gathered}
a(\partial\xi)  = \int_{\partial\xi} f 
\qquad
a^{*}(\partial\xi) = \int_{\partial\xi} f^{*}
\\[1ex]
\phi(\xi)\rightarrow \phi(\xi)+ a(\partial\xi)
\qquad
\phi^{*}(\xi)\rightarrow \phi^{*}(\xi)+ a^{*}(\partial\xi)
\\[1ex]
V_{p,p^{*}}(\xi)\rightarrow V_{p,p^{*}}(\xi)
\;e^{ipa(\partial\xi)+ip^{*}a^{*}(\partial\xi)}
\end{gathered}
\en
Fix an $(n{-}2)$-boundary $\partial\xi_{0}$ and consider
the abelian subgroup of $\cD^{\sing}_{n-1}(M)$
\eq
\cD^{\sing}_{n-1}(M)_{\Integers\partial\xi_{0}}
=\left\{\xi\colon
\:\partial\xi \in \Integers\partial\xi_{0}
\right\}
\;\subset
\cD^{\sing}_{n-1}(M)
\en
On $\cD^{\sing}_{n-1}(M)_{\Integers\partial\xi_{0}}$
the gauge 
symmetries act as a global $U(1){\times} U(1)$ generated by the
two numbers $a(\partial\xi_{0})$ and $a^{*}(\partial\xi_{0})$.

\newsec{5}
Calculus is needed on 
$\cD^{\sing}_{n-1}(M)_{\Integers\partial\xi_{0}}$
to continue the analogy with the 2d gaussian model.
Go to the metric completion,
writing it $Q=\cD^{\integral}_{n-1}(M)_{\Integers\partial\xi_{0}}$.
Geometric Measure Theory provides
a construction of currents in any such complete 
metric space \cite{AmbrosioKirchheim},
providing the spaces $\cD^{\integral}_{j}(Q)$ of integral $j$-currents in $Q$.
Define the $j$-forms on $Q$ as the real duals of the currents
and the exterior derivative as the dual of the boundary operator,
\eq
\Omega_{j}(Q) = \Hom(\cD^{\integral}_{j}(Q),\Reals)
\qquad
d\omega(\eta) = \omega(\partial\eta)
\en
The infinitesimal
$j$-simplices  generate $\cD^{\integral}_{j}(Q)$,
so the tangent bundle $TQ$ can be defined as the set of infinitesimal 
1-simplices in $Q$.  The 1-forms then become the sections of the dual 
cotangent bundle $T^{*}Q$.

The
equivalences of simplices $\Delta^{j}\times \Delta^{n-1}\cong \Delta^{j+n-1}$
give natural maps
\eq
\Pi_{j,n-1}\colon\,\cD^{\integral}_{j}(Q)\rightarrow 
\cD^{\integral}_{j+n-1}(M)
\qquad
\partial \Pi_{j,n-1} = \Pi_{j-1,n-1} \partial
\en
The map $\Pi_{1,n-1}$
identifies each tangent space $ T_{\xi}Q$ 
with a certain subspace $\cV_{n}\subset\cD^{\distr}_{n}(M)$.
That $*\cV_{n} = \cV_{n}$
is a crucial technical point
whose demonstration uses
the flat metric completion.
Then $*$ acts on each tangent space $ T_{\xi}Q$.
The forms $F$, $*F$, $A$, $A^{*}$ on $M$ pull back to
$j$, $*j$, $\phi$, $\phi^{*}$ on $Q$,
\eq
\begin{gathered}
j=\Pi_{1,n-1}^{*}F
\quad
{*}j = \Pi_{1,n-1}^{*}(*F)
\qquad
\phi = \Pi_{0,n-1}^{*} A
\quad
\phi^{*} = \Pi_{0,n-1}^{*} A^{*}
\\[1ex]
d\phi = j
\qquad
d\phi^{*} = {*}j
\end{gathered}
\en
So there is the classical 2d gaussian model on each of the spaces $Q$,
except that $*^{2}=1$ for $n$ even,
while $*^{2}=-1$ in 2d.
Define
\eq
J=\epsilon_{n}*
\qquad
\epsilon_{n}^{2} =(-1)^{n-1}
\qquad
J^{2}=-1
\en
$J$ is imaginary when $n$ is even, so the currents
have to be complexified
in order that $J$ act on the tangent spaces $T_{\xi}Q$,
\eq
Q= \cD^{\integral}_{n-1}(M)_{\Integers\partial\xi_{0}}
\oplus i \partial \cD^{\integral}_{n}(M)
\en
Now, for all $n$, on each of the 
spaces $Q$ there is a 2d gaussian model
written in terms of 
the chiral fields
\eq
d\phi_{\pm}= j_{\pm}
\qquad
j_{\pm}=\frac12(1\pm i^{-1}J)j
\en

\newsec{6}
Quantization of a free field theory is
expressed by the Schwinger-Dyson equation on the 2-point functions.
In the 2d gaussian model, the chiral fields are (anti-)holomorphic.
The S-D equation on $\langle \bar\phi_{\pm}\,j_{\pm}\rangle$
is the Cauchy-Riemann equation
\eq
\frac{\partial}{\partial \bar z} \;\frac{1}{z-z'} = \pi
\delta^{2}(z-z')
\en
which is the foundation for complex analysis on Riemann 
surfaces.
The 2d gaussian model would have led to complex analysis on Riemann surfaces
had that not existed already.
For the free $n$-form in $2n$ dimensions,
the S-D equation 
has an expression containing no explicit mention of $n$,
\eq
\langle
\int_{\bar \xi_{1}} \bar A_{\alpha}
\int_{ \xi_{2}} d F_{\beta}
\rangle
=
2\pi i c_{\alpha\beta}
\IM{\bar\xi_{1}}{\xi_{2}}
\qquad
c_{\alpha\beta}= -c_{\beta\alpha}
\quad
c_{+-}=1
\label{eq:SD}
\en
The lhs is the 2-pt function
$
\langle
\bar A_{\pm}(x)
\,d F_{\pm}(x')
\rangle
$
smeared against the
$(n{-}1)$-current $\bar\xi_{1}$
and the $(n{+}1)$-current $\xi_{2}$.
The rhs is a slight modification of the intersection number,
which is
nonzero only if $k_{1}+k_{2}=2n$,
\eq
I_{M}(\xi_{1},\xi_{2})
=
\int_{M}
\,
\frac1{k_{1}!\;k_{2}!}
\xi_{1}^{\mu_{1}\cdots}(x)
\,
\xi_{2}^{\nu_{1}\cdots\mu_{k_{1}}}(x)
{\epsilon_{\mu_{1}\cdots \mu_{k_{1}}\nu_{1}\cdots\nu_{k_{2}}}}(x)
\;d^{2n}x
\en
The modification is such that $\IM{\bar\xi_{1}}{\xi_{2}}$
has properties independent of $n$,
\eq
\IM{\bar\xi_{1}}{\xi_{2}} = 
\epsilon_{n,k_{2}-n} I_{M}(\bar\xi_{1}, \xi_{2})
\qquad
\epsilon_{n,m} = (-1)^{n m+m(m+1)/2} \,\epsilon_{n}^{-1}
\en
\begin{align}
&\text{$\IM{\bar\xi_{1}}{\xi_{2}}$ is skew-hermitian.}
\\
&\text{$\IM{\overbar{\partial\xi_{1}}}{\xi_{2}} 
= -\IM{\bar\xi_{1}}{\partial\xi_{2}}$}
\\
&\text{$\IM{\bar\xi_{1}}{J\xi_{2}}$ on $n$-currents is hermitian and 
positive definite}
\end{align}
Pulled back to $Q$, the S-D equation of the free $n$-form cft is
\eq
\langle
\int_{\bar\eta_{1}} \bar\phi_{\alpha}
\int_{ \eta_{2}} d j_{\beta}
\rangle
=
2\pi i c_{\alpha\beta}
I_{Q}\langle\bar \eta_{1},\eta_{2}\rangle
\label{eq:SDQ}
\en
The rhs is $\IM{\bar\xi_{1}}{\xi_{2}}$
pulled back to a skew-hermitian form on currents in $Q$
\eq
I_{Q}\langle\bar \eta_{1},\eta_{2}\rangle =  \IM{\Pi_{j_{1},n-1}\bar\eta_{1}}{\Pi_{j_{2},n-1}\eta_{2}}
\en
which
is nonzero only if $(j_{1}+n-1)+(j_{2}+n-1) = 2n$,
which is $j_{1}+j_{2}=2$,
just like the intersection number of currents in a 2-manifold.
The S-D equation (\ref{eq:SDQ}) on $Q$ is formally analogous to the S-D equation
(\ref{eq:SD}) of the 2d gaussian model on a Riemann surface,
which is the Cauchy-Riemann equation.

\newsec{7}
The free $n$-form cft on $M$ has now become the
2d gaussian model on each of the metric spaces $Q$.  
Moreover, each  $Q$ has the structure
needed to write the Cauchy-Riemann equation.
This is taken to be the defining structure of a {\it quasi Riemann
surface}.
The $Q=\cD^{\integral}_{\Integers\partial\xi_{0}}$
are  the quasi Riemann surfaces.
They are the fibers of a bundle of quasi Riemann surfaces
\eq
\begin{gathered}
\cQ(M) \rightarrow \PBM\\[1ex]
\PBM =\{\text{maximal infinite cyclic subgroups }\Integers\partial\xi_{0}
\subset \partial\cD^{\integral}_{n-1}(M)\}
\end{gathered}
\en
On each fiber $Q$
there is a 2d gaussian model
with its global $U(1){\times} U(1)$ symmetry group, collectively
comprising a local gauge symmetry over $\PBM$.

\newsec{8}
All of the constructions of 2d cft are based on
the Cauchy-Riemann equation and on the 2d gaussian model.
So there is the prospect of carrying out those constructions on each 
of the fibers $Q$
to obtain, for 
every 2d cft, a new cft of defects in $M$.
The 2d cft on each fiber $Q$ will be ambiguous up to its global 2d symmetry 
group.
The collection of global symmetry groups on the fibers
will forms a local gauge symmetry group over $\PBM$.

\newsec{9}
There are many basic problems to be worked on:
opportunities to leverage 2d qft to develop a new technology of qft
in $2n$ dimensions.
Some are the following.

Complex analysis on quasi Riemann surfaces needs to be 
developed in analogy with ordinary Riemann surfaces.

Conjecturally, every quasi Riemann surface $Q$ is 
isomorphic to $\cD^{\integral}_{0}(\Sigma)$ for $\Sigma$ the 2d conformal 
space with the same Jacobian as $Q$,
the Jacobian being the complex torus made from the homology in the 
middle dimension.

The conjectured isomorphism
would allow constructing a 2d cft on each $Q$ by lifting an 
ordinary 2d cft from $\Sigma$ to $\cD^{\integral}_{0}(\Sigma)$,
which is a purely 2d problem,
then to each $Q=\cD^{\integral}(M)_{\Integers\partial\xi_{0}}$ via 
one of the conjectured isomorphisms.
As an example of the first step,
the 2d gaussian model is to be lifted
by extending the renormalization of the vertex operators $V_{p,\bar p}(\xi)$
from singular 0-currents $\xi$ to integral 0-currents.

If the conjecture is true, there will be some universal 
objects to study.  The automorphism group $\Aut(Q)$ will encode 
information about all the conformal groups of the conformal manifolds $M$
and about the global symmetry groups of all 2d cfts.
There will be a universal homogeneous bundle of quasi Riemann surfaces 
with structure group $\Aut(Q)$
in which all the bundles $\cQ(M)\rightarrow\PBM$ are embedded.

There should be a large collection of structure preserving maps from 
the complex disk $\Disk_{1}$ into $Q$. Meromorphic 
functions on $Q$ will pull back to ordinary meromorphic functions on $\Disk_{1}$.  
The local structure of a cft on $Q$ will be 
expressed as a collection of ordinary 2d cfts on each of these {\it 
local quasi holomorphic 
curves}, each with its radial quantization, pair of Virasoro 
algebras, and operator product expansion.
Explicit
constructions of local quasi holomorphic curves are needed, say for $M=S^{2n}$.

The local gauge symmetry in the bundle of quasi Riemann surfaces
needs a space-time interpretation.
What, for example,  is the space-time interpretation
of the local $SU(2){\times} SU(2)$ symmetry over $\PBM$ corresponding to the global 
$SU(2){\times} SU(2)$ at the self-dual 
point $R=1$ of the 2d gaussian model?

It should be possible to imitate on the quasi Riemann surfaces the 
usual constructions of 2d cft such as 
orbifolding and perturbation theory.

Tiny defects look like points in $M$, so fields $\Phi(\xi)$ 
when restricted to the small $\xi$ in $Q$
will give ordinary local quantum fields on $M$.
Will these form new local qfts in $2n$-dimensions?

\vspace{3ex}
\providecommand{\href}[2]{#2}
\begingroup
\raggedright
\renewcommand{\section}[2]{}%

\endgroup

\end{document}